\documentclass[conference]{IEEEtran}
\IEEEoverridecommandlockouts
\usepackage{cite}
\usepackage{amsmath,amssymb,amsfonts}
\usepackage{algorithmic}
\usepackage{graphicx}
\usepackage{textcomp}
\usepackage{xcolor}
\usepackage{multirow}
\usepackage{url}
\usepackage{csquotes}

\newcommand{\squeezeup}{\vspace{-2.5mm}}
\newcommand{\squeezeupsmall}{\vspace{-0.5mm}}
\def\BibTeX{{\rm B\kern-.  05em{\sc i\kern-.025em b}\kern-.08em
    T\kern-.1667em\lower.7ex\hbox{E}\kern-.125emX}}
\begin{document}
\title{RCA Copilot: Transforming Network Data into Actionable Insights via Large Language Models
\squeezeup

}
\author{

    \IEEEauthorblockN{Alexander Shan}
    \IEEEauthorblockA{
    \textit{Stanford University} \\
    U.S.A. \\
    azshan@cs.stanford.edu}
    \\[0.1ex] 
    \IEEEauthorblockN{Tarun Banka}
    \IEEEauthorblockA{\textit{Juniper Networks} \\
    Sunnyvale, U.S.A. \\
    tbanka@juniper.net}
    \and
    \IEEEauthorblockN{Jasleen Kaur}
    \IEEEauthorblockA{\textit{Juniper Networks} \\
    Sunnyvale, U.S.A. \\
    jkaur@juniper.net}
    \\[0.1ex] 
    
    \IEEEauthorblockN{Raj Yavatkar}
    \IEEEauthorblockA{\textit{Juniper Networks} \\
    Sunnyvale, U.S.A. \\
    ryavatkar@juniper.net}
    \and
    \IEEEauthorblockN{Rahul Singh}
    \IEEEauthorblockA{\textit{Juniper Networks} \\
    Bangalore, India \\
    rahulsingh@juniper.net}
    \\[0.1ex]
    \IEEEauthorblockN{T. Sridhar}
    \IEEEauthorblockA{\textit{Juniper Networks} \\
    Sunnyvale, U.S.A. \\
    tsridhar@juniper.net}
}
\maketitle
\squeezeup
\squeezeup
\begin{abstract}
Ensuring the reliability and availability of complex networked services demands effective root cause analysis (RCA) across cloud environments, data centers, and on-premises networks. Traditional RCA methods, which involve manual inspection of data sources such as logs and telemetry data, are often time-consuming and challenging for on-call engineers. While statistical inference methods have been employed to estimate the causality of network events, these approaches alone are similarly challenging and suffer from a lack of interpretability, making it difficult for engineers to understand the predictions made by black-box models. In this paper, we present RCACopilot, an advanced on-call system that combines statistical tests and large language model (LLM) reasoning to automate RCA across various network environments. RCACopilot gathers and synthesizes critical runtime diagnostic information, predicts the root cause of incidents, provides a clear explanatory narrative, and offers targeted action steps for engineers to resolve the issues. By utilizing LLM reasoning techniques and retrieval, RCACopilot delivers accurate and practical support for operators.
\end{abstract}

\squeezeup
\squeezeup
\section{INTRODUCTION}
\squeezeupsmall
As reliance on multi cloud platforms, data centers, and hybrid on-premises networks grows, ensuring consistent service reliability and availability across these complex environments become critical. In large scale networks, unexpected service interruptions or performance degradation can severely impact customer satisfaction. Currently, the diagnosis of such incidents predominantly relies on manual investigation or the use of specialized data filtering tools. Given the increasing scale and complexity of contemporary networking systems, operator intervention alone is insufficient for the effective and timely resolution of incidents.

Root cause analysis is a critical component of the incident management lifecycle, essential for identifying the underlying causes of incidents \cite{wu2020microrca}. Through root cause analysis, engineers can pinpoint the fundamental issues that triggered the incident, allowing for the implementation of corrective measures to prevent future occurrences. This process is vital for effective incident resolution, enhancing system reliability, and improving overall incident response mechanisms \cite{wang2024comprehensive}.

Despite the promising performance of large language models (LLMs) in incident diagnosis \cite{roy2024exploring} tasks—particularly when fine-tuned on incident data—they encounter several challenges when applied to root cause analysis. First, the current fine-tuning approaches assume that the model can learn all the intricate details of past incidents. However, it is well-established that LLMs are prone to hallucinations, often producing distorted or exaggerated information, as they cannot reliably recall specific details from the training data. Additionally, fine-tuning large LLMs is associated with considerable costs and may be infeasible for state-of-the-art models with extremely large parameter counts, such as GPT-4 \cite{achiam2023gpt}. Lastly, outdated knowledge from past training data becomes obsolete in the face of emerging information, motivating continuous learning methods instead. Continuous fine-tuning is required to keep the model updated with the latest knowledge, but this presents significant challenges in maintaining the model's capacity to incorporate new information effectively.

These concerns drive the adoption of a non-finetuning approach for automating root cause analysis. This approach must harness domain knowledge similarly to finetuning, but without the associated drawbacks. In this work, we develop a few shot learning based system to eliminate the need for finetuning. Our method wields the dual strengths of statistical RCA methods and LLM reasoning abilities. In conjunction with estimating the causality of network events, we incorporate a retrieval mechanism that selects past RCA cases as in-context exemplars, helping the LLM apply the appropriate diagnostic methods and domain expertise. Additionally, we introduce advanced prompting techniques to improve consistency and enhance the reasoning depth of model responses. 

\section{RELATED WORK}
\subsection{Statistical Root Cause Analysis}
\label{statistical rca}
Automatic root cause analysis has been investigated prior to the advent of large language models, leveraging statistical causality tests based on time series data. Prior studies \cite{kobayashi2017mining} implemented a statistical method to construct network event causality graphs by mining time series data and applying the PC algorithm. Pham, Ha, and Zhang \cite{pham2024root} reference a suite of statistical methods surrounding estimating Granger causality to construct a causality graph of events. Then, the Pagerank algorithm can be used to rank the most likely root causes. Modern methods produce a ranked list of nodes which can either represent events across a time series, or a network component (e.g. a switch). These approaches are both fast and technically sound; in our implementations of statistical tests, we find that the true root cause is ranked in the top five of our predictions. However, there exist limitations to statistical approaches when it comes to the utility to operators/on-call engineers. For instance, a ranked list only exists as a suggestion of probable causes--it remains left to the operator to determine the proper solution. Furthermore, without any reasoning for its decision, the process of validating the model's root cause prediction is left to the on-call engineer, which is often time-consuming and mitigates the impact of automating RCA.

\subsection{Large Language Model-based Root Cause Analysis}
Large language models such as GPT-4 and Llama 3 have been trained on a vast text corpus comprising a significant portion of human generated writing on the internet. This training data enables these models to perform a variety of downstream tasks. As a result of LLMs' emergent capabilities for reasoning and complex problem solving, researchers have attempted to leverage LLMs for root cause analysis.

For example, Ahmed et al. \cite{ahmed2023recommending} demonstrated the first capability for LLMs to diagnose the root cause of network incidents by generating a diagnosis from the title and description of an incident. Ahmed finetuned GPT 3.5 on a small set of incident titles and descriptions along with corresponding root causes. Chen et al. \cite{chen2024automatic} expanded upon Ahmed's work, creating a retrieval based system for generating root cause predictions based on past cases. The notable improvement upon their work was the ingestion of more extensive data sources and generating reasoning chains about different possibilities of root causes. Chen et al. \cite{chen2024automatic} explored root cause prediction with LLMs without the use of finetuning. In many downstream tasks, finetuning is used on an out-of-the-box LLM in order to augment the model with domain knowledge it would otherwise lack access to. However, by using relevant in-context examples, Zhang et al. \cite{duan2024scalable} demonstrated that competitive performance can be achieved without finetuning. Additionally, models exhibited lower levels of hallucination and qualitatively stronger reasoning chains while exploring hypotheses.

Areas of exploration in this field we aim to expand upon are the synthesis of multiple modeling techniques, such as retrieval, tool usage, and prompt engineering. Individual papers have explored these concepts in isolation, but no work has shown the result of a combination. Besides in-context learning, prompt engineering techniques have flown under the radar, including self-consistency, chain-of-thought prompting, prompt chaining, and knowledge generation. We investigate whether it is possible to leverage the advantages of prompt engineering and higher quality inputs to outperform finetuned LLMs.
\squeezeup
\squeezeupsmall
\section{DATASET SOURCING}
\subsection{Experimental data}
\label{Experimental data}
\begin{figure}[ht]
    \centering
    \includegraphics[width=1.0\linewidth]{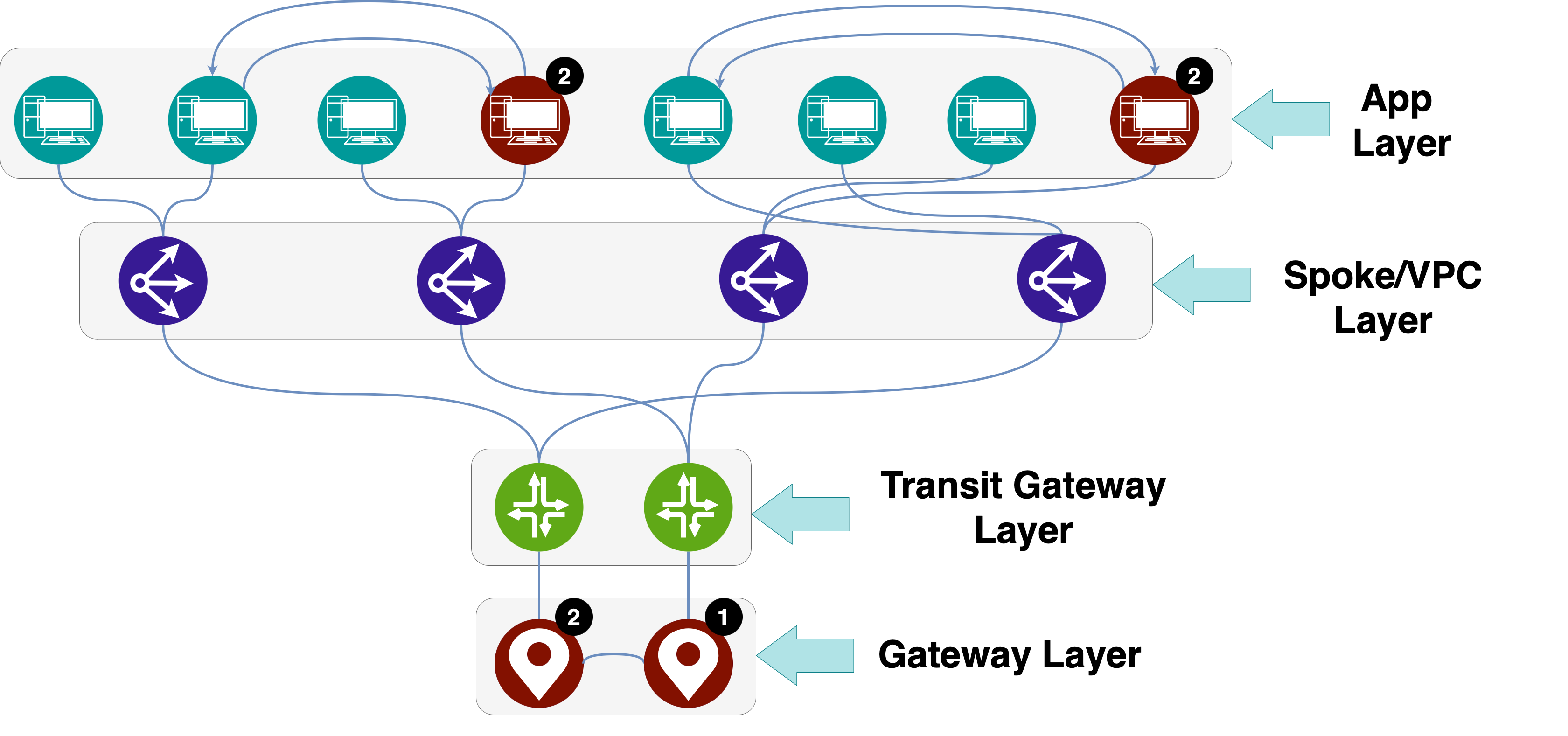}
    \caption{Vista Hybrid and Multi-cloud Network Topology}
    \label{fig:Network Fault}
\end{figure}

\begin{figure}[h]
    \centering
    \includegraphics[width=1.0\linewidth]{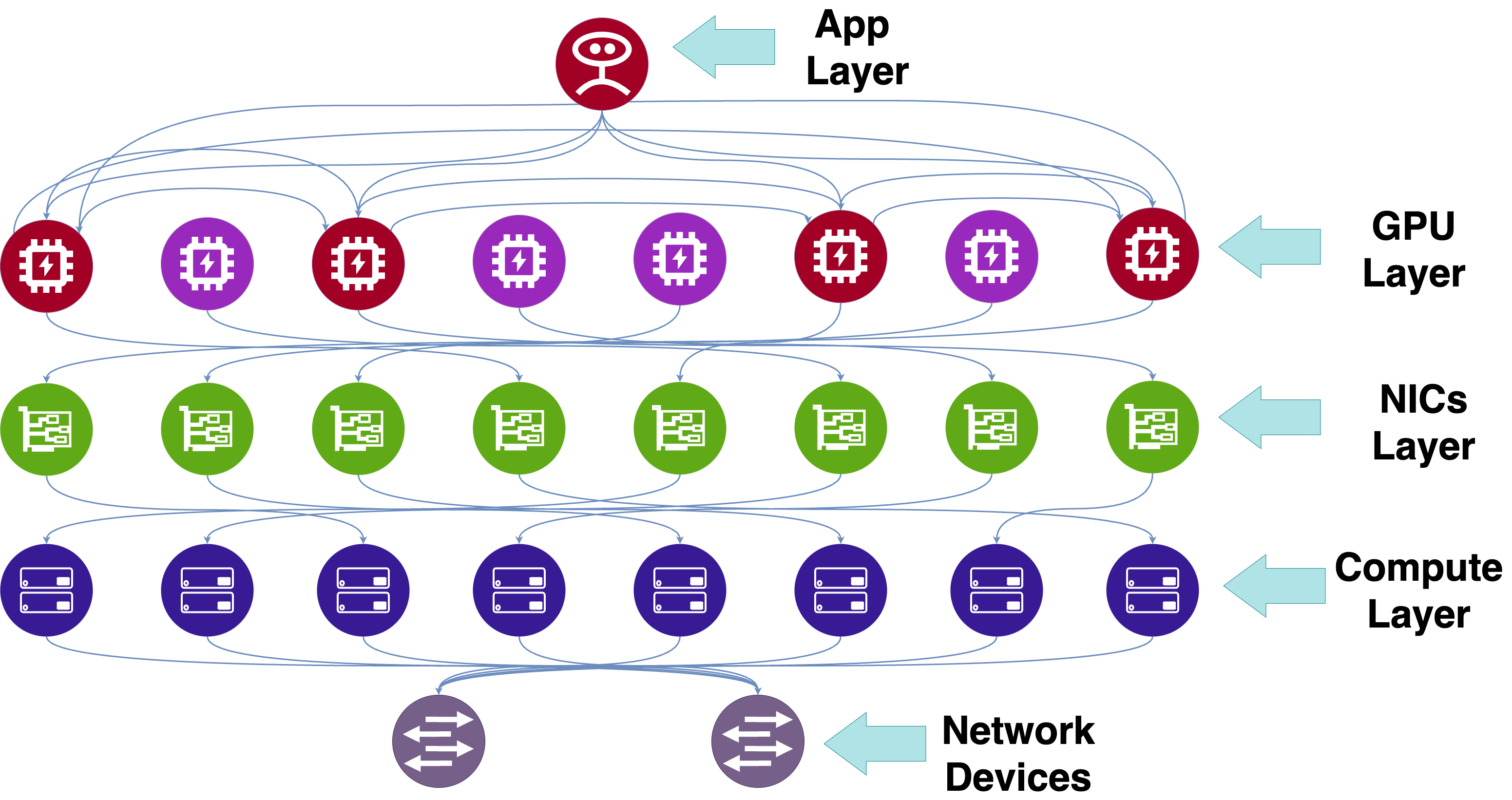}
    \caption{On-Prem AIML Workload Topology}
    \label{fig:AIMLAssurance}
\end{figure}

We configured are two different network topologies, each with 5 and 3 unique fault scenarios respectively. Figure \ref{fig:Network Fault} represents the first network topology comprises of a large-scale hybrid and multicloud network, with four layers: Application layers that possess application endpoints or subnets, Spokes that are the group of app nodes, Transit Gateways (TGW) which is an AWS construct and Gateways or Smart session routers that are application aware. In each layer, there are multiple nodes that are connected with each other. In our setup, we inject faults at different time periods to reflect realistic incident scenarios that may occur in customer environments. To construct our dataset, we collect samples of the network topology states across different time periods. 
\squeezeup
\squeezeup
%FEEDBACK
Figure \ref{fig:AIMLAssurance} illustrates the second network topology, depicting a real-world network infrastructure designed to support distributed AI/ML workloads in a datacenter. This topology comprises of 5 layers, Application: AI applications with training and inference workloads, GPU: Graphics Processing Units for AI apps, NICs: Network Interface Cards for connectivity, Compute: where application instances are running and Network devices: Network switches in the fabric.
\squeezeupsmall
\squeezeup
\squeezeup
\\
A state consists of two pieces of information: 

1. \textbf{Snapshot of the Topology graph:} This captures all the nodes in the topology, their connectivity to other nodes, and their telemetry data. More information about this diagnostic information will be described in \ref{Diagnostic data collection}.

2. \textbf{Statistical RCA result:} This is a JSON object representing an oracle perspective on the root causes of application layer anomalies. More information about this item will be explained in Section \ref{Approach: stat rca}.

For each state, our dataset keeps track of the ground truth diagnosis and resolution action steps. Since we injected faults at controlled intervals, we assigned each state a proper diagnosis and wrote a corresponding reasoning chain to deduce the answer and accompanying resolution action steps and call it the gold diagnosis. During evaluation, we query RCACopilot on the input states and compare its answer to the gold diagnosis and action steps.
\squeezeupsmall
\squeezeupsmall
\subsection{Diagnostic Information}

\label{Diagnostic data collection}
As input to RCACopilot, we include a set of diagnostic information required to identify the proper root cause. In contrast to Zhang and Chen's work \cite{duan2024scalable}, we do not include stacktraces or code retrieved from related support tickets in our diagnostic information. Instead, we include metric data for each node and the Network graph topology directly as JSON-formatted data. This gives the LLM knowledge about different anomalies at each network layer, as well as the connectivity between nodes at different network layers. Additionally, operator domain knowledge can be directly injected into the input prompt, such as known relationships between metrics. 

% FEEDBACK original modiWe also include a text description of application layer anomalies.%
We also include a brief text description in the prompt explaining what the anomalous metric means in a real network scenario.
These reflect the downstream impacts of the incident and serve as a starting point for generating root cause hypotheses. We generate the description by prompting an LLM to summarize the application layer metrics and highlight any anomalies, which are marked in the metric data. 

There are some data sources for each network topology that are unique to our approach. Specifically, device telemetry data (e.g. CPU utilization on a router, or latency metric on application) serves as an important source of truth for RCACopilot. Moreover, device logs, network flow data, and traces all serve as further clues for RCACopilot to form, validate, and discredit different hypotheses.

Finally, we make use of statistical RCA results as input to RCACopilot in a similar way to section \ref{statistical rca}. In particular, we collect time series data on device metrics as inputs to the Granger causality test. The produced graph is then filtered using the PageRank algorithm to produce a ranking of nodes and metrics that are likely to indicate the root cause of application anomalies. This information serves as an oracle to the LLM, filtering the input data to highlight key areas of interest to form hypotheses from. We format this information as a system data health report, with the top $K$ ranked nodes highlighted in the prompt ($K=5$) in our uses.

\section{METHODOLOGY}
\subsection{Architecture Overview}
\begin{figure}[h]
\squeezeup
    \includegraphics[scale=0.29]{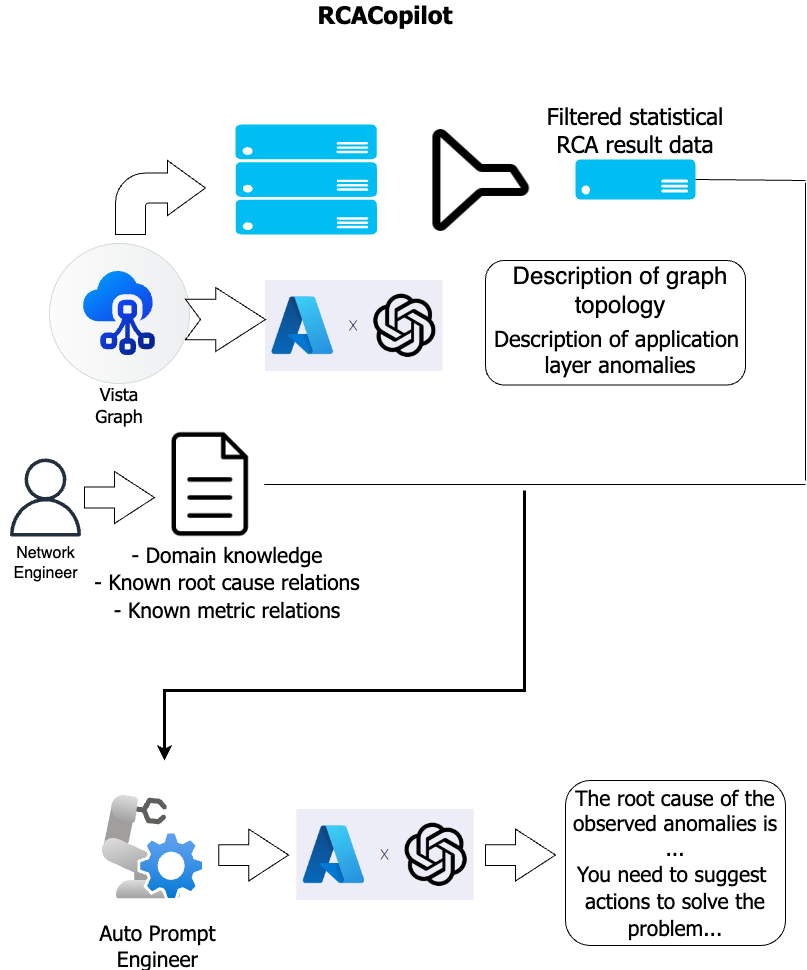}
    \caption{Architecture Diagram of RCA Copilot}
    \label{fig:RCA_Copilot_Architecture_Diagram}
\squeezeup
\squeezeupsmall
\end{figure}

Figure \ref{fig:RCA_Copilot_Architecture_Diagram} outlines the design of RCACopilot. When an anomaly is detected at the application layer (e.g. application node experiences latency), the workflow is triggered. First, diagnostic data sources are queried to reconstruct the state of the Network Graph Topology (node metrics, connectivity between nodes at each layer, etc.). Some data sources, such as those representing graph topology and application layer symptoms, must be summarized by an intermediate LLM call. Other data sources such as the statistical RCA ranked list are presented in their unprocessed JSON forms.

After all of the data are fetched, we compile it into a lengthy prompt to a powerful LLM (e.g. GPT-4) and instruct it to form a hypothesis for the root cause of all observed application issues. Additionally, for each issue, the LLM is instructed to form a set of specific action steps that will resolve the incident(s). In this section, we cover the details of the techniques we use to enhance model performance.
\squeezeupsmall
\subsection{Statistical RCA}
\label{Approach: stat rca}
Our statistical RCA rank list serves as a filter on input data, prioritizing the most likely root causes of observed issues. This allows RCACopilot to function as an interpretive layer on top of these tests, offering deeper reasoning capabilities beyond statistical results. 

To compute statistical RCA we create a causality graph. Firstly, we extract anomalous nodes from the topology graph, and run time series correlation tests, granger causality \cite{granger}  to find out cause and effect relationships. With the help of the correlation results, we determine directions between the anomalous nodes and construct a graph. Then we run Pearson correlation to compute the strength of these relationships and assign weights to the edges in the causal graph. Lastly we run the page rank algorithm \cite{pagerank} to find the top K root cause from the causal graph. 

\subsection{Retrieving In-Context Examples}
Zhang et al. \cite{pham2024root} show that providing LLMs with relevant few shot, or in-context examples, dramatically improves root cause prediction accuracy. However, choosing relevant examples to provide the model is an important and difficult step. We define relevance to constitute two components; first, examples with similar topologies produce stronger signals. Second, the relevance of the incident issue(s) and corresponding diagnosis drives the strongest improvement. This is because similar incident issues or diagnoses often share resolution strategies and hypotheses, leading to them being more powerful in-context examples. Thus, we built a retrieval system to select these desired examples.

To construct our retrieval dataset, we use the same approach as in \ref{Experimental data}, but on a different set of states: we collect eight graph topology states with faults that we have the ground truth causes for. These states have paired gold diagnoses and action steps written in text. Then, we use an embedding model to generate a vector representation for the diagnostic information. The embedding, root cause explanation, and action steps for a given topology are indexed into a vector database, e.g. FAISS \cite{douze2024faiss}, to enable efficient similarity search. When a novel incident arises, we use its diagnostic information as a query to find relevant incidents, diagnoses, and action steps based on the retrieval index. The extracted examples are then prepended into the prompt of the LLM. Finally, we utilize the LLM, such as GPT-4, to generate the root cause based on the new incident description and all the provided in-context examples. 
%\begin{figure}[ht]
%    \centering
%    \includegraphics[width=1.0\linewidth]{imgs/retrieval_diagram.jpeg}
%    \caption{Retrieval Structure for RCA Copilot. \cite{b2}}
%    \label{fig:Retrieval Structure for RCA Copilot}
%\end{figure}

Via this retrieval system, RCA Copilot can use domain knowledge relevant to specific use cases and incident types. As more retrieval examples are added, RCA Copilot can apply specific facts or approaches learned in one case to a new one. This system gives RCA Copilot stronger performance across time, strengthening the value of the system as it oversees a topology for longer.

\subsection{Eliciting Reasoning}
We employ an ensemble of prompt engineering techniques to accomplish two goals; to improve the quality of reasoning chains and to reduce hallucinations. For example, we use chain of thought prompting \cite{wei2022chain} \cite{kojima2022large} to encourage RCACopilot to form and explore various hypotheses. We also use in-context learning beyond the hypothesis formation step; when forming action steps, we provide pairs of hypotheses and their corresponding resolution steps. Additionally, we apply a mixture of agents \cite{wang2024mixture} to our LLM inference calls in order to capture the consensus of multiple hypotheses. 

\section{EXPERIMENTS}
\subsection{Experimental Setup}
For our setup, we collect eight samples of graph topology states across different times. 
In our setup, there are 2 different network topologies (Figure \ref{fig:Network Fault} and \ref{fig:AIMLAssurance}) , each with 5 and 3 unique fault scenarios respectively. %FEEDBACK These faults are generated using various tools, such as iperf for high traffic generation, stress-ng for CPU load increase, and TGW route table resets to induce packet drops. %
Here is a summary of each of the usescases:
\begin{enumerate}
\renewcommand{\labelenumi}{\theenumi.}
\setlength{\leftmargin}{0pt}
\item \textbf{High App Bandwidth}: Heavy burst of traffic is injected between two application endpoint due to excessive data transfer  
\item \textbf{High App Latency}: This scenario involves latency at the application layer, where only one node experiences delays without anomalies in other layers. One of the application nodes suffers from anomalous latency reporting high time to first data packet.  
\item \textbf{Over-utilization of the GPU}: High GPU utilization across nodes causes premature iteration completion in the application layer, leading to potential inaccuracies or incomplete data processing.
\item \textbf{Nic Ack Timeout Error}: Significant packet acknowledgment ACK timeout errors in the NICs layer cause communication delays, leading to incomplete data processing and premature iteration completion in the AIApps node.
\item \textbf{TGW Blackhole}: A blackhole in the Transit Gateway route disrupts traffic flow, causing packet drops and triggering repeated TCP retransmissions between app endpoints. 
\item \textbf{Gateway Packet Loss}: Packet loss in the gateway layer increases network delays, causing extended time to receive the first data packet at the application layer. 
\item \textbf{Gateway Resource Contention}: Excessive CPU usage on gateway nodes delays packet processing, resulting in network slowdowns and increased TCP retransmissions between apps. 
\item \textbf{Switch Congestion}: Network congestion in the NICs and Switches layers, along with overloaded GPUs at full utilization, limits application processing capacity, causing delays and congestion notification (CNP) being sent in the NICs layer. 
\end{enumerate}

In each of the experimental usecases, we conduct an evaluation as explained in \ref{evaluation_section} demonstrating RCA Copilot abilities to find the root cause in a network topology configured with injected faults. 
\squeezeupsmall
\squeezeupsmall
\subsection{Evaluation Criteria}
\label{evaluation_section}
To evaluate the effectiveness of RCA Copilot, we have devised an automatic evaluation mechanism to score the responses of RCA Copilot on specific network topology states. We define the output of RCA Copilot inference to be composed of the entire text generated by the LLM containing its chain of thought leading to the diagnosis, the diagnosis itself, and the corresponding action steps at each layer.

To evaluate the output, we must measure the similarity between two texts (or sets of texts). In our case, we evaluate the similarity between the generated text from RCA Copilot and a ground truth text diagnosis and action steps. We use two methods to compute similarity scores; first, we compute the BERTScore \cite{zhang2019bertscore} between the true explanation and the copilot-generated explanation. Second, we use a sentence transformer such as S-BERT \cite{reimers2019sentence} to take the embedding over the true and predicted text. Then, we compute the cosine similarity between the embeddings to produce a semantic similarity metric. 

We compute the results in two types of environments, Few Shot and Zero Shot. In few-shot environment, the model (GPT-4 in our usecase) is given a prompt containing a few examples to guide it in completing a task. These examples serve as context, helping the model generate more accurate outputs by following the patterns in the prompt. In zero-shot environment, the model is provided only with an instruction or question, without any examples. The model relies solely on its pre-trained knowledge to generate a relevant response.

\section{RESULTS}
The following Table \ref{experiment1} summarizes the results of various use cases evaluated in a few-shot environment using BERTScore and S-BERT metrics. The BERTScore measures precision (P), recall (R), and F1-scores to assess the semantic similarity of predictions, while S-BERT captures contextual similarity through sentence embeddings. The results indicate high performance across most use cases, with TGW Blackhole achieving the highest scores (0.95), reflecting strong alignment with expected outcomes. In contrast, Switch Congestion shows relatively lower scores (~0.81), suggesting room for improvement. Overall, RCA Copilot demonstrates effective generalization with minimal examples, especially for network-related tasks.
\squeezeup

\begin{table}[h]
\caption{RCA Copilot Performance in Few Shot Environment}
\label{experiment1}
\begin{tabular}{|c|c|ccc|c|}
\hline
\multirow{2}{*}{\textbf{SNo}} & \multirow{2}{*}{\textbf{Usecase}}                                               & \multicolumn{3}{c|}{\textbf{BertScore}}                                         & \multirow{2}{*}{\textbf{\begin{tabular}[c]{@{}c@{}}S-Bert \\ Score\end{tabular}}} \\ \cline{3-5}
                              &                                                                                 & \multicolumn{1}{c|}{\textbf{F1}} & \multicolumn{1}{c|}{\textbf{P}} & \textbf{R} &                                                                                   \\ \hline
\textbf{1}                    & \textbf{\begin{tabular}[c]{@{}c@{}}High App \\ Bandwidth\end{tabular}}          & \multicolumn{1}{c|}{0.86}        & \multicolumn{1}{c|}{0.86}       & 0.84       & 0.89                                                                              \\ \hline
\textbf{2}                    & \textbf{\begin{tabular}[c]{@{}c@{}}High App \\ Latency\end{tabular}}            & \multicolumn{1}{c|}{0.88}        & \multicolumn{1}{c|}{0.89}       & 0.86       & 0.92                                                                              \\ \hline
\textbf{3}                    & \textbf{\begin{tabular}[c]{@{}c@{}}High GPU\\  Utilization\end{tabular}}        & \multicolumn{1}{c|}{0.83}        & \multicolumn{1}{c|}{0.83}       & 0.85       & 0.90                                                                              \\ \hline
\textbf{4}                    & \textbf{\begin{tabular}[c]{@{}c@{}}Nic ACK \\ Timeout Error\end{tabular}}       & \multicolumn{1}{c|}{0.86}        & \multicolumn{1}{c|}{0.85}       & 0.87       & 0.92                                                                              \\ \hline
\textbf{5}                    & \textbf{\begin{tabular}[c]{@{}c@{}}TGW \\ Blackhole\end{tabular}}               & \multicolumn{1}{c|}{0.95}        & \multicolumn{1}{c|}{0.95}       & 0.95       & 0.93\\ \hline
\textbf{6}                    & \textbf{\begin{tabular}[c]{@{}c@{}}Gateway \\ Packet Loss\end{tabular}}         & \multicolumn{1}{c|}{0.91}        & \multicolumn{1}{c|}{0.91}       & 0.89       & 0.88 \\ \hline
\textbf{7}                    & \textbf{\begin{tabular}[c]{@{}c@{}}Gateway \\ Resource Contention\end{tabular}} & \multicolumn{1}{c|}{0.91}        & \multicolumn{1}{c|}{0.91}       & 0.89       & 0.92 \\ \hline
\textbf{8}                    & \textbf{\begin{tabular}[c]{@{}c@{}}Switch \\ Congestion\end{tabular}}           & \multicolumn{1}{c|}{0.81}        & \multicolumn{1}{c|}{0.80}       & 0.80       & 0.93                                                                              \\ \hline
\end{tabular}
\squeezeup
\end{table}

Table \ref{experiment2} presents the results of various use cases evaluated in a zero-shot environment using BERTScore and S-BERT metrics. Compared to the few-shot setting, the scores are generally lower, reflecting the challenge of performing tasks without prior examples. Notably, BERTScore shows more consistent and reliable performance across use cases, with higher precision and recall values, making it a better indicator of semantic similarity. BERTScore emerges as a more true indicator of semantic relevance, as it reflects meaningful alignment with the underlying task outcomes. In contrast, S-BERT primarily captures structural or contextual similarity, which can be less precise in zero-shot scenarios.  For example, TGW Blackhole and Gateway Resource Contention show higher BERTScores (0.66), suggesting better alignment with expected behavior when no specific task guidance is provided through examples in the zero shot environment.
\begin{table}[h]
\squeezeup
\caption{RCA Copilot Performance in Zero Shot Environment}
\label{experiment2}
\begin{tabular}{|c|c|ccc|c|}
\hline
\multirow{2}{*}{\textbf{SNo}} & \multirow{2}{*}{\textbf{Usecase}}                                               & \multicolumn{3}{c|}{\textbf{BertScore}}                                         & \multirow{2}{*}{\textbf{\begin{tabular}[c]{@{}c@{}}S-Bert \\ Score\end{tabular}}} \\ \cline{3-5} &      & \multicolumn{1}{c|}{\textbf{F1}} & \multicolumn{1}{c|}{\textbf{P}} & \textbf{R} &     \\ \hline
\textbf{1}                    & \textbf{\begin{tabular}[c]{@{}c@{}}High App \\ Bandwidth\end{tabular}}          & \multicolumn{1}{c|}{0.52}        & \multicolumn{1}{c|}{0.53}       & 0.54       & 0.78 \\ \hline
\textbf{2}                    & \textbf{\begin{tabular}[c]{@{}c@{}}High App \\ Latency\end{tabular}}            & \multicolumn{1}{c|}{0.52}        & \multicolumn{1}{c|}{0.52}       & 0.52       & 0.83 \\ \hline
\textbf{3}                    & \textbf{\begin{tabular}[c]{@{}c@{}}High GPU\\  Utilization\end{tabular}}        & \multicolumn{1}{c|}{0.64}        & \multicolumn{1}{c|}{0.61}       & 0.66       & 0.88 \\ \hline
\textbf{4}                    & \textbf{\begin{tabular}[c]{@{}c@{}}Nic ACK \\ Timeout Error\end{tabular}}       & \multicolumn{1}{c|}{0.50}        & \multicolumn{1}{c|}{0.50}       & 0.50       & 0.82 \\ \hline
\textbf{5}                    & \textbf{\begin{tabular}[c]{@{}c@{}}TGW \\ Blackhole\end{tabular}}               & \multicolumn{1}{c|}{0.66}        & \multicolumn{1}{c|}{0.64}       & 0.68       & 0.84 \\ \hline
\textbf{6}                    & \textbf{\begin{tabular}[c]{@{}c@{}}Gateway \\ Packet Loss\end{tabular}}         & \multicolumn{1}{c|}{0.48}        & \multicolumn{1}{c|}{0.48}       & 0.49       & 0.77 \\ \hline
\textbf{7}                    & \textbf{\begin{tabular}[c]{@{}c@{}}Gateway \\ Resource Contention\end{tabular}} & \multicolumn{1}{c|}{0.66}        & \multicolumn{1}{c|}{0.66}       & 0.67       & 0.89 \\ \hline
\textbf{8}                    & \textbf{\begin{tabular}[c]{@{}c@{}}Switch \\ Congestion\end{tabular}}           & \multicolumn{1}{c|}{0.51}        & \multicolumn{1}{c|}{0.53}       & 0.49       & 0.78                                                                              \\ \hline
\end{tabular}
\squeezeup
\squeezeupsmall
\end{table}

\section{DISCUSSION}
We observe a few notable trends in RCA Copilot's abilities after our internal testing on various topologies and fault injection scenarios. In this section, we describe the most important observations with respect to performance and hence, future work. 
\subsection{Use Case Definition}
\squeezeupsmall
\squeezeupsmall
\label{Gateways latency use case}

Consider the usecase \textbf{Gateway Resource Contention}: we injected a fault in the Gateway Layer, using the stress-ng tool, we overloaded the CPU which resulted in an increase in the CPU Utilization metric in the Gateway node. This induced resource exhaustion at the gateway, which in turn impacted the Application Layer’s performance, causing delays and degraded service quality. Figure \ref{fig:Network Fault} displays the graphical representation of a cloud network; the afflicted Gateway layer nodes at the bottom of the figure, as well as the impacted application endpoints, are highlighted in red. Table \ref{table_3} describes the results from the statistical RCA analysis, determining probable root cause rank 1 to be cpu utilization in the SSR routers and rank 2 to be the application latency metrics. 
\begin{table}[h]
\squeezeup
\caption{Statistical RCA Results}
\squeezeup
\label{table_3}
\begin{tabular}{|c|c|c|c|}
\hline
\textbf{Rank} & \textbf{Layer} & \textbf{Node}                 & \textbf{Metric}             \\ \hline
\textbf{1}    & Gateways       & \begin{tabular}[c]{@{}c@{}}VistaDev-aws-\\ us-west-2\end{tabular}    & total\_cpu\_utilization                                                                     \\ \hline
\textbf{2}    & Application    & \begin{tabular}[c]{@{}c@{}}Sausalito-spoke-\\ us-east-2\end{tabular} & \begin{tabular}[c]{@{}c@{}}applications\_time\_to\_\\ first\_data\_packet\_avg\end{tabular} \\ \hline
\textbf{2}    & Application    & \begin{tabular}[c]{@{}c@{}}Sausalito-spoke-\\ us-east-2\end{tabular} & \begin{tabular}[c]{@{}c@{}}applications\_ack\_\\ round\_trip\_forward\_avg\end{tabular}     \\ \hline
\end{tabular}
\squeezeup
\end{table}
In response to this application latency, an operator can enable RCA Copilot for on-demand analysis. RCA Copilot will compute summary using inputs from the topology graph information in Figure \ref{fig:Network Fault} Statistical results in Table \ref{table_3}, and few shot learning examples with specific prompts guiding the model as described in IV-C.
%{imgs/RCA_Copilot_UI_GatewayCPU_Alt.png}
%\begin{figure}[h]
%    \centering
%    \includegraphics[width=1.0\linewidth]
    %{imgs/RCA_Copilot_UI_GatewayCPU_Alt.png}
%    \caption{RCA Copilot Results summary}
%    \label{fig:RCACopilot diagnosis}
%\squeezeup
%\squeezeup
%\end{figure}

\begin{table}[h]
\caption{RCA Copilot Diagnosis}
\label{RCACopilot diagnosis}
\begin{tabular}{|l|l|}
\hline
\textbf{Symptom}                                                                                                                     & \textbf{High Latency} in the Application Layer                  \\ \hline
\textbf{\begin{tabular}[c]{@{}l@{}}Root cause\\ hypothesis\end{tabular}}                                                             & \begin{tabular}[c]{@{}l@{}}The root cause of the high latency is likely \\ due to the \textbf{high CPU utilization} on the \\ Gateway node VistaDev-aws-us-west-2. \\ The high CPU utilization can cause delays \\ in processing packets, leading to increased \\ acknowledgment round trip times \\ and overall latency.\end{tabular} \\ \hline
\multirow{2}{*}{\textbf{\begin{tabular}[c]{@{}l@{}}Action Steps\\ on  Gateway \\ Layer Node \\ VistaDev-aws-us-west-2\end{tabular}}} & \begin{tabular}[c]{@{}l@{}}1. \textbf{Reduce the CPU load} on the \\ Gateways node VistaDev-aws-us-west-2 \\ as the high CPU utilization on this node is \\ likely causing delays in packet\\ processing, leading to increased latency.\end{tabular}                    \\ \cline{2-2} 
& \begin{tabular}[c]{@{}l@{}}2. \textbf{Implement load balancing} across the\\ Gateways  nodes to distribute the \\ processing load more evenly\end{tabular}     \\ \hline
\end{tabular}
\squeezeup
\squeezeup
\end{table}
Table \ref{RCACopilot diagnosis} shows the summary of the RCA Copilot diagnosis. The symptom of high latency is highlighted and the root cause is localized to the Gateways layer. The RCA Copilot identifies that Application node being affected and High Latency as the symptom. It provides a concise hypothesis, and indicates the high data packet latency is because of the high CPU Utilization issue in one of the gateway layer nodes. Then it recommends actions steps that can be taken on the gateway layer node. By performing these action steps, we can mitigate anomalies in the networking stack. In this case there was only one symptom; for complex network topologies with nodes continuously degrading, each symptom will have its own hypothesis---a two to three sentence summary of the predicted root cause and its effect on the applications. 
As seen in the hypothesis and action steps, RCA Copilot is capable of reasoning about cross-layer problems. It also suggests specific actions referencing particular nodes, layers, and metrics in the topology graph that it has observed in its prompt context.  

\subsection{Effectiveness}

RCA Copilot demonstrated human engineer-level analysis in a set of networking stack and fault injection scenarios from the experiments. Future work will evaluate across a broader range of network deployment scenarios and various fault situations at different scales. 
RCA Copilot showcased basic reasoning in environments where the few shot examples lack relevance to the current symptom. In one fault scenario, the only in-context examples were from Gateway layer issues, but when injecting Transit Gateway layer packet blackholes, RCA Copilot correctly determined the root cause with a perfect chain of thought. This indicates that for RCA Copilot a small set of few shot examples were good enough to root cause unseen scenarios, and hence our approach is scalable. The chain of thought structure demonstrated in the examples can be enough to boost performance, as seen in \cite{pham2024root},%FEEDBACK 
without the additional overhead of resources to fine-tune the model. 
\subsection{Prompt Engineering}
Since we interact with the GPT-4 model through API, the only improvements we can make to model outputs is through improving RCA Copilot's prompts. We observe that the most influential factor on RCA the performance is the relevance and quality of the few shot exemplars present in the context window of the model at inference time. Therefore, we will prioritize enhancing our RAG architecture to ensure that selected examples are relevant for diagnosis, especially within smaller context windows. We will also explore other LLMs for root cause analysis in the future. 

\section{CONCLUSION}
\squeezeupsmall
In this work, we present RCA Copilot, an LLM-based system for on-demand cross-layer root cause analysis used in incident management. We detail how RCA Copilot finds and synthesizes runtime diagnostic information, predicts the cause(s) of incidents, and offers a targeted set of actions for operators to mitigate the problem(s). RCA Copilot relies on a few-shot approach to inference, where the exemplars are retrieved from a corpus of known incidents and resolutions. We feature a variety of prompting techniques to enhance the reasoning capabilities of the model and reduce hallucinations. Based on our experiments, RCA Copilot boasts strong accuracy and coherence when tested across network topologies. We also introduce a novel evaluation framework to measure the performance of automatic RCA systems based on semantic similarity and classification scores. Finally, we offer an analysis of our work pertaining to the benefits and limitations of our approach. 
\squeezeupsmall

\bibliographystyle{ieeetr}
\bibliography{references}

\vspace{12pt}
\end{document}